# Optical Transparency Using Interference Between Two Modes of a Cavity

J.D. Franson and S.M. Hendrickson
*University of Maryland, Baltimore County, Baltimore, MD 21250 USA*

In electromagnetically-induced transparency (EIT), the scattering rate of a probe beam is greatly reduced due to destructive interference between two dressed atomic states produced by a strong laser beam. Here we show that a similar reduction in the single-photon scattering rate can be achieved by tuning a probe beam to be halfway between the resonant frequencies of two modes of a cavity. This technique is expected to be useful in enhancing two-photon absorption while reducing losses due to single-photon scattering.

In electromagnetically-induced transparency (EIT) [1,2], an atomic medium with a large scattering rate at one frequency can be made transparent by applying a laser beam at a different frequency. The reduction in the scattering rate is due to destructive interference between the scattering amplitudes from two dressed atomic states that are created by the laser beam. Here we show that a similar effect can be obtained without the need for a laser beam if the frequency of an incident beam of light is tuned to be halfway between two of the resonant modes of a cavity. This allows the two-photon absorption rate to be enhanced while simultaneously reducing the loss due to single-photon scattering. Aside from its fundamental interest, this effect is expected to be of practical use in the design of Zeno logic gates based on two-photon absorption [3].

The system of interest here is illustrated in Fig. 1. An optical waveguide containing a small number of incident photons is coupled to a ring resonator, which in turn is coupled to $N_A$ three-level atoms via its evanescent field. It will be assumed that there is no direct coupling of the photons in the waveguide to the atoms, so that no loss can occur unless a photon has been coupled into the resonator. The atoms are located near the coupling to the waveguide in order to ensure that the matrix elements are the same for adjacent resonator modes. Our goal is to achieve strong absorption when two photons are present in the waveguide with little or no loss due to scattering when only a single photon is present. It will be found that the loss due to single-photon scattering can be eliminated if the frequency of the incident photons is tuned to be halfway between two adjacent cavity modes.

Ring resonators can have large intrinsic quality factors (Q values), in which case there would be very little response to an incident field tuned between two of the resonant modes. But if the coupling between the waveguide and the ring resonator is very strong, the Q can be reduced (spoiled) to the point that the linewidth is comparable to the free spectral range of the cavity.

Johnsson et al. [4] considered the use of EIT in two-photon absorption based on a conventional approach in which a strong laser beam is applied to a double-lambda atomic system. Opatrny and Welsch [5] have previously discussed an enhanced form of EIT using the coupled modes of two different cavities and a strong pump beam (laser). Classical interference effects involving two different cavities (but no atoms) that are somewhat analogous to EIT have also been investigated [6-8]. Our approach has the potential advantage of avoiding any background counts due to the laser beam, which may be important when dealing with single-photon signals. In addition, confining the photons to the small mode volume of the ring resonator enhances the magnitude of the two-photon absorption and is well-suited for quantum computing applications [3].

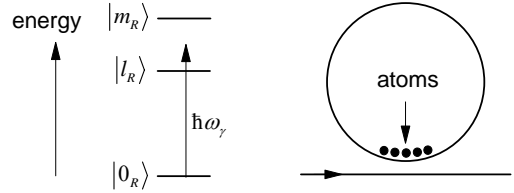

Fig. 1. Reduction in the single-photon scattering rate by tuning the photons in a waveguide halfway between the resonant frequencies of two adjacent resonator modes $|l_R\rangle$ and $|m_R\rangle$. The single-photon loss is greatly reduced due to destructive interference between the probability amplitudes to excite an atom, while strong two-photon absorption can still occur.

The energy level diagram for the system of interest here is illustrated in Fig. 2. The ring resonator has resonant modes labeled $l_R$ that correspond to $l_R \lambda = C$, where $C$ is the circumference of the ring and $\lambda$ is the wavelength. Neglecting dispersion, the energy $E(l_R)$ of a single photon in the state $|l_R\rangle$ is given by $E(l_R) = l_R \hbar \omega_0$, where $\hbar$ is Planck's constant divided by $2\pi$ and $\omega_0$ is the angular frequency of the fundamental mode. The energy $\hbar \omega_\gamma$ of the incident photons can be written as $\hbar \omega_\gamma = \bar{E} + \delta$, where $\delta$ is an adjustable parameter and $\bar{E} = (l_R \hbar\omega_0 + m_R \hbar\omega_0)/2$ is the average energy of two adjacent modes $l_R$ and $m_R = l_R + 1$, as shown in Fig. 2. The resonator state $|l_R, m_R\rangle$ with one photon in modes $l_R$ and $m_R$ has an energy of $2\bar{E}$, and the state $|2m_R\rangle$ with two photons in mode $m_R$ will also play an important role. For simplicity, the other modes of the ring resonator are not shown in Fig 2.

The ground state and first two excited states of the atoms will be denoted by $|0_A\rangle$, $|1_A\rangle$, and $|2_A\rangle$, respectively. The energy $E_1$ of $|1_A\rangle$ is assumed to be detuned by $\Delta_1$ from $\bar{E}$, while the energy $E_2$ of $|2_A\rangle$ is detuned by $\Delta_2$ from $2\bar{E}$.

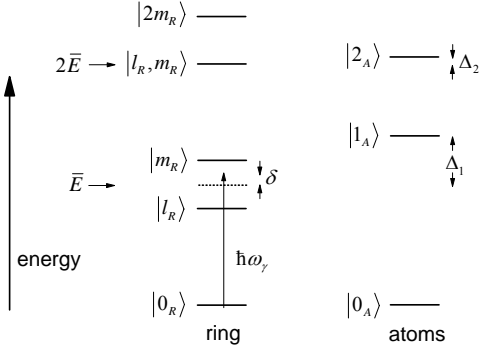

Fig. 2. Energy level diagram for the ring resonator and three-level atoms. Quantum interference between the resonator modes $|l_R\rangle$ and $|m_R\rangle$ eliminates single-photon scattering by the atoms in analogy with EIT, while strong two-photon absorption can still occur via intermediate states such as $|2m_R\rangle$.

All of the atoms are confined to a sufficiently small region that $\exp\{i[k(l_R) - k(m_R)]x\} \doteq 1$, where $x$ is the distance around the ring. In that case the exponential phase factors can be neglected and the Hamiltonian for a single atom in the dipole approximation is given [5] by

$$\hat{H} = \hat{a}_W^\dagger \hat{a}_W \hbar\omega_\gamma + \sum_l \hat{a}_l^\dagger \hat{a}_l \hbar\omega_l + E_1 \hat{\sigma}_{01}^\dagger \hat{\sigma}_{01} + E_2 \hat{\sigma}_{12}^\dagger \hat{\sigma}_{12}$$
$$+ M_W \sum_l \hat{a}_l^\dagger \hat{a}_W + M_1 \sum_l \hat{\sigma}_{01}^\dagger \hat{a}_l + M_2 \sum_l \hat{\sigma}_{12}^\dagger \hat{a}_l + h.c. \quad (1)$$

Here $\hat{a}_W^\dagger$ and $\hat{a}_l^\dagger$ create photons in the waveguide and ring resonator, respectively, $\hat{\sigma}_{01}$ produces a transition from $|1_A\rangle$ to $|0_A\rangle$, and $\hat{\sigma}_{12}$ produces a transition from $|2_A\rangle$ to $|1_A\rangle$. The value of the coefficient $M_W$ is determined by the coupling (evanescent field) between the waveguide and the ring resonator, while the atomic matrix elements are given by $M_1 = <\vec{d}_1 \cdot \vec{E}>$ and $M_2 = <\vec{d}_2 \cdot \vec{E}>$, where $\vec{d}_1$ and $\vec{d}_2$ are the corresponding dipole moments and $\vec{E}$ is the electric field in the resonator [9].

We will first consider the case in which the atomic density is sufficiently low that perturbation theory can be used, and then return later to consider the implications of higher atomic densities. To lowest order in perturbation theory, the rate of single-photon scattering due to the presence of the atoms is given [10] by

$$R_1 = 2N_A \frac{M_{eff}^2}{(\hbar\omega_\gamma - E_1)^2 + (\hbar\gamma_1)^2} \gamma_1. \quad (2)$$

Here $\gamma_1$ is the half-width of state $|1_A\rangle$ due to spontaneous emission and collision broadening and $M_{eff}$ is the effective matrix element from second-order perturbation theory [11]. There are two probability amplitudes leading to the excitation of state $|1_A\rangle$ corresponding to the virtual excitation of resonator states $|l_R\rangle$ or $|m_R\rangle$. The effective matrix element for this process is thus the sum of two terms

$$M_{eff} = \left( \frac{M_1 M_W}{\delta - \hbar\omega_0/2} + \frac{M_1 M_W}{\delta + \hbar\omega_0/2} \right) \quad (3)$$

where $\delta \pm \hbar\omega_0/2$ is the detuning in the two intermediate states and we assume that $\delta \ll \hbar\omega_0$.

Combining Eqs. (2) and (3) gives

$$R_1 = 2N_A \left( \frac{\gamma_1}{\Delta_1^2 + (\hbar\gamma_1)^2} \right) \left( \frac{M_1 M_W}{\delta - \hbar\omega_0/2} + \frac{M_1 M_W}{\delta + \hbar\omega_0/2} \right)^2 \quad (4)$$

where we have used $\delta \ll \Delta_1$. The same results were also obtained using a more lengthy density matrix calculation, which gives a small correction for the intrinsic linewidth of the resonator which is negligible for the high-Q cavities of interest here [12].

The rate of single-photon scattering as a function of the tuning parameter $\delta$ is indicated by the solid line in Fig. 3 in arbitrary units. (The numerical values as a function of atomic density will be discussed below.) It can be seen that the scattering rate goes to zero for $\delta = 0$ due to the cancellation between the two intermediate states. Since the cancellation occurs between two existing states, this effect is more analogous to a Fano resonance [13] than EIT. The scattering rate is further reduced by the large value of $\Delta_1$ regardless of the cancellation of amplitudes. The scattering amplitudes from the modes not shown in Fig. 3 are smaller in magnitude and they also cancel out in pairs.

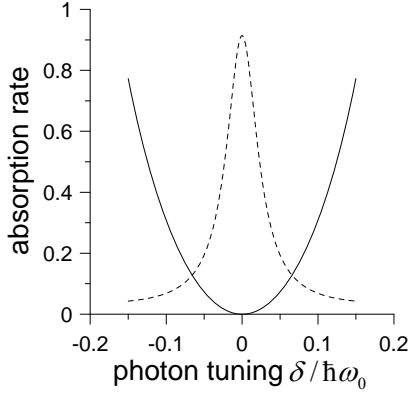

Fig. 3. Single-photon scattering rate (solid line) and two-photon absorption rate (dashed line) as a function of the photon detuning $\delta/\hbar\omega_0$ (in arbitrary units).

Our goal is to achieve a large rate of two-photon absorption, and the energy levels are such that two photons in the waveguide could be resonantly absorbed into the state the state $|l_R, m_R\rangle$ of the resonator for $\delta = 0$. But to lowest order in perturbation theory, that process is also suppressed by the cancellation of probability amplitudes, since the intermediate states for two-photon absorption of that kind are the same as in Eq. (3).

This problem can be avoided if we include the possibility of two-photon absorption via the resonator state $|2m_R\rangle$, for example. There is only one probability amplitude for this transition since both photons are absorbed into mode $m$ of the ring resonator, and no cancellation of probability amplitudes occurs. Two-photon absorption can occur in fourth order perturbation theory via the transition

$$|2_W\rangle \to |1_W, m_R\rangle \to |2m_R\rangle \to |m_R, 1_A\rangle \to |2_A\rangle. \quad (5)$$

Here $|2_W\rangle$ is the initial state with two photons in the waveguide, $|1_W, m_R\rangle$ corresponds to one photon in the waveguide and one in resonator mode $m$, etc. The probability amplitude $A_2$ for this transition is given [11] by

$$A_2 = \frac{M_2}{(2\delta - \Delta_2)} \frac{\sqrt{2}M_1}{\Delta_1} \frac{\sqrt{2}M_W}{(2\delta - \hbar\omega_0)} \frac{M_W}{(\delta - \hbar\omega_0/2)} \quad (6)$$

where we have assumed $\delta \ll \Delta_1$. There is a similar fourth-order probability amplitude $A_2'$ to excite the upper atomic state via resonator states $|l\rangle$ and $|2l\rangle$:

$$A_2' = \frac{M_2}{(2\delta - \Delta_2)} \frac{\sqrt{2}M_1}{\Delta_1} \frac{\sqrt{2}M_W}{(2\delta + \hbar\omega_0)} \frac{M_W}{(\delta + \hbar\omega_0/2)}. \quad (7)$$

It is important to note that the detunings in the first two virtual states are both positive or both negative for either process (for $|\delta| < \hbar\omega_0/2$), so that these two amplitudes have the same sign and constructively interfere. Adding these two amplitudes gives a two-photon absorption rate from Eq. (2) of

$$R_2 = 8N_A \frac{M_1^2 M_2^2 M_W^4 \gamma_2}{\Delta_1^2 [(2\delta)^2 + (\hbar\gamma_2)^2]}$$

$$\times \left( \frac{1}{(2\delta - \hbar\omega_0)} \frac{1}{(\delta - \hbar\omega_0/2)} + \frac{1}{(2\delta + \hbar\omega_0)} \frac{1}{(\delta + \hbar\omega_0/2)} \right)^2. \quad (8)$$

Here $\gamma_2$ is the width of the upper atomic level due to collisions and we have taken $\Delta_2 = 0$.

Eq. (8) includes only the contribution from the two intermediate states with the smallest detunings, which give most of the contribution to the probability amplitude. The total probability amplitude was determined by numerically summing the contributions from all possible intermediate states $|l_i\rangle$ and $|l_j, l_k\rangle$, which gave a two-photon absorption rate that is 52% larger than that from Eq. (8). The dashed line in Fig. 3 shows the total two-photon absorption rate as a function of the photon tuning parameter $\delta$ for the case of $\hbar\gamma_2 = 0.05 \times \hbar\omega_0$. It can be seen that the two-photon absorption rate is a maximum at $\delta = 0$ where the single-photon scattering rate is zero.

The coupling between state $|2m_R\rangle$ and the states $|1_A\rangle$ for each of the $N_A$ atoms will be too large for standard perturbation theory to be valid if the density of atoms is sufficiently large [14]. In that case, the rate of two-photon absorption can be determined by first calculating the exact eigenstates of the Hamiltonian $\hat{H}_0$ defined to include the diagonal elements of $\hat{H}$ as well as the $M_1$ term that couples those states. Perturbation theory can then be performed using those eigenstates as a basis with a perturbation Hamiltonian $\hat{H}'$ that includes the remaining $M_W$ and $M_2$ terms, which are independent of the atomic density. The probability amplitude for an atom to be in state $|1_A\rangle$ will be denoted by $c_{1A}$, while $c_{2m}$ will denote the probability amplitude for state $|2m_R\rangle$. Schrodinger's equation for $\hat{H}_0$ gives

$$i\hbar \dot{c}_{2m} = (2\bar{E} + \hbar\omega_0)c_{2m} + \sqrt{2}N_A M_1 c_{1A}$$
$$i\hbar \dot{c}_{1A} = (2\bar{E} + \hbar\omega_0/2 + \Delta_1)c_{1A} + \sqrt{2}M_1 c_{2m} \quad (9)$$

If we make a change of variables to $c'_{2m} \equiv \sqrt{N_A} c_{2m}$ and $M'_1 \equiv \sqrt{N_A} M_1$, then Eq. (9) can be rewritten as

$$i\hbar \frac{d}{dt}\begin{bmatrix} c'_{2m} \\ c_{1A} \end{bmatrix} = \begin{bmatrix} (2\bar{E} + \hbar\omega_0) & \sqrt{2}M'_1 \\ \sqrt{2}M'_1 & (2\bar{E} + \hbar\omega_0/2 + \Delta_1) \end{bmatrix}\begin{bmatrix} c'_{2m} \\ c_{1A} \end{bmatrix} \quad (10)$$

which is equivalent to an effective two-level system. The eigenstates of Eq. (10) were found analytically and combined with the remaining states ($|0_R\rangle$ and $|2_A\rangle$) to form a new set of basis states for the perturbation calculation in $M_W$ and $M_2$.

In order to estimate the magnitude of these effects, we considered a simple example of a ring resonator formed from a fused silica fiber with diameter $d$ in the shape of a ring (toroid) of diameter $D \gg d$. This provides an approximation to the whispering gallery modes of toroidal resonators [15]. In the limit $D \gg d$, the electric field is that of a straight fiber for which there is an exact solution to Maxwell's equations [16]. This allows a determination of the matrix elements $M_1 = \langle \vec{d}_1 \cdot \vec{E} \rangle$ and $M_2 = \langle \vec{d}_2 \cdot \vec{E} \rangle$. For this example, we assumed that $d = 0.35 \ \mu m$ and $D = 50 \ \mu m$, which corresponds to a mode spacing of $\omega_0/2\pi = 1.8 \times 10^{12}$ Hz and an effective mode volume $V_m = 7.6 \times 10^{-17}$ m$^3$ [9]. The coupling between the waveguide and the cavity was taken to be $M_W = \hbar\omega_0/3$.

It was also assumed that the resonator is surrounded by rubidium vapor at density $\rho$ with $N_A = \rho V_m$. The levels $|1_A\rangle$ and $|2_A\rangle$ were taken to be the usual rubidium $S \rightarrow P \rightarrow D$ transition [17] used in precise wavelength measurements with $\gamma_1 = \gamma_2 \doteq \pi \times 10^8$ sec$^{-1}$. For this series of transitions, the two-photon absorption resonance is at 778 nm and the detuning $\Delta_1$ corresponds to a wavelength difference of 2.1 nm.

The two-photon absorption rate calculated in this way is shown in Fig. 4. At densities up to $\sim 10^{15}$ cm$^{-1}$, the straight line nature of the plot is identical to Eq. (8), while $R_2$ saturates at a value of $4.7 \times 10^9$ sec$^{-1}$ for larger densities. Roughly speaking, the saturation is due to the fact that a single photon can be absorbed in a virtual process by at most one atom at a time regardless of how many atoms are present. For comparison purposes, the single-photon scattering rate at $\rho = 10^{15}$ cm$^{-3}$ is $R_1 = 2.3 \times 10^7$ sec$^{-1}$ for $\delta/\hbar\omega_0 = 0.2$, which can be effectively eliminated using the quantum interference effects discussed above. The intrinsic losses in a high-Q ring resonator can be below $10^5$ sec$^{-1}$, which suggests that two-photon absorption rates that are orders of magnitude larger than the single-photon loss should be experimentally achievable using these techniques.

In summary, tuning between two resonant modes of a cavity with a small mode volume can result in two-photon absorption rates that are much larger than the single-photon loss. These techniques are expected to be of practical use in the implementation of Zeno logic gates [3] and single-photon sources based on two-photon absorption. Although we have not discussed it here, similar techniques could also be used to produce nonlinear phase shifts with low single-photon loss.

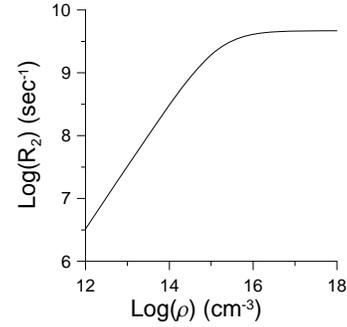

Fig. 4. Plot of the two-photon absorption rate $R_2$ as a function of the density $\rho$ of rubidium vapor. The saturation at high densities is due to the fact that a single photon can be absorbed in a virtual process by at most one atom at a time.

We would like to acknowledge discussions with Almut Beige, Bryan Jacobs, and Todd Pittman. This work was supported by DTO, ARO, and IR&D funding.